# Structural Transformations of Carbon and Boron Nitride Nanoscrolls at High Impact Collisions

C. F. Woellner[a], L. D. Machado[b], P. A. S. Autreto[c], J. M. de Sousa[a,d,e], and D. S. Galvao[a*]

[a.] *Applied Physics Department and Center for Computational Engineering & Sciences, State University of Campinas, Campinas-SP, 13083-970, Brazil.*
[b.] *Departamento de Física Teórica e Experimental, Universidade Federal do Rio Grande do Norte, Natal-RN 59072-970, Brazil.*
[c.] *Universidade Federal do ABC, Santo André-SP, 09210-580, Brazil.*
[d.] *Departamento de Física, Universidade Federal do Piauí, Teresina-PI, 64049-550, Brazil*
[e.] *Departamento de Física, Universidade Federal do Ceará, P.O. Box 6030, CEP 60455-900, Fortaleza-CE, Brazil.*

The behavior of nanostructures under high strain-rate conditions has been object of theoretical and experimental investigations in recent years. For instance, it has been shown that carbon and boron nitride nanotubes can be unzipped into nanoribbons at high velocity impacts. However, the response of many nanostructures to high strain-rate conditions is still not completely understood. In this work we have investigated through fully atomistic reactive (ReaxFF) molecular dynamics (MD) simulations the mechanical behavior of carbon (CNS) and boron nitride nanoscrolls (BNS) colliding against solid targets at high velocities,. CNS (BNS) nanoscrolls are graphene (boron nitride) membranes rolled up into papyrus-like structures. Their open-ended topology leads to unique properties not found in close-ended analogues, such as nanotubes. Our results show that the collision products are mainly determined by impact velocities and by two impact angles, which define the position of the scroll (i) axis and (ii) open edge relative to the target. Our MD results showed that for appropriate velocities and orientations large-scale deformations and nanoscroll fracture can occur. We also observed unscrolling (scrolls going back to quasi-planar membranes), scroll unzipping into nanoribbons, and significant reconstruction due to breaking and/or formation of new chemical bonds. For particular edge orientations and velocities, conversion from open to close-ended topology is also possible, due to the fusion of nanoscroll walls.

## Introduction

The mechanical properties of nanomaterials have been object of intense investigations in the last years. Carbon nanostructures are of particular interest, as the elastic modulus (over 1 TPa[1,2]) and tensile strength (over 100 GPa[1,3]) of nanotubes and graphene are the highest ever measured. Hexagonal boron nitride (h-BN) nanostructures have also received considerable attention, as they present comparable modulus[4] and strength[5], but superior chemical stability[6] and torsional stiffness[7]. These characterizations are usually carried out in the low strain-rate regime, where inertia effects are not very important[8]. However, for ballistic impacts these effects cannot be neglected. The high strength and low weight of carbon and h-BN nanomaterials make them ideal candidates for ballistic protection applications. However, the rapidly increasing strain characteristics of impact events can lead to stress localization and fractures[9], thus possibly affecting their mechanical performance. Recently, many works have been dedicated to the mechanical characterization of nanostructures in the high strain-rate regime[10,11,12,13,14,15], including the observation of record-breaking energy absorption capabilities for graphene-based materials[9].

In order to generate high strain-rate conditions, the typical procedure is to shoot projectiles against stationary nanostructures or, vice-versa, shooting nanomaterials against stationary targets. Recently, carbon[16] and h-BN[17] nanotubes were shot as projectiles against aluminum targets at hypervelocity (5-7 km/s). Corresponding molecular dynamics (MD) simulations[16,17] revealed that the critical parameters to determine the resulting structures after impact are the relative nanotube/target orientation and impact velocity values. For certain conditions, unzipping of nanotubes into nanoribbons and the formation of nanodiamonds were also observed[18]. These results suggest that hypervelocity impacts can be used as an effective tool to break and/or generate new structures. In principle, this approach can be applied to a large class of nanomaterials, such as nanoscrolls.

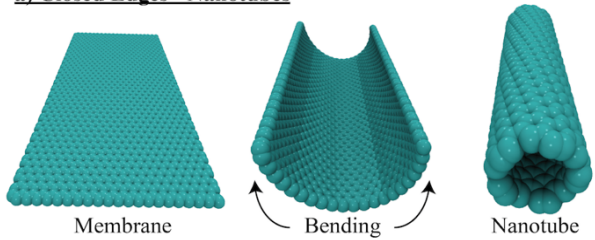
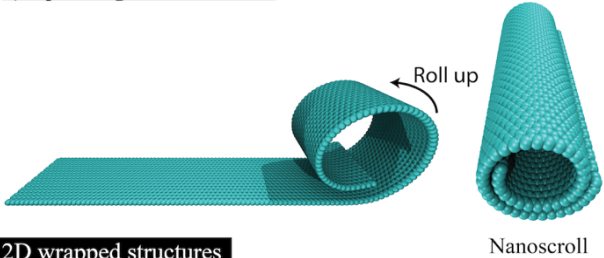

Figure 1: Rolled-up carbon nanostructure topologies: (a) closed edges as nanotubes and; (b) open edges as nanoscrolls

Carbon (CNS) and boron nitride (BNS) nanoscrolls are nanostructures obtained from rolling-up graphene and hexagonal BN sheets into a papyrus-like shape[19,20] – see figure 1. Interestingly, the final scrolled structures can have lower energy than the initial planar configurations[21,22], due to van der Waals interactions at the overlapping nanoscroll regions. Both CNS[23,24,25,26] and BNS[26,27,28] have been already synthesized, and their open-ended nature can be exploited in many applications. For instance, when compared to multi-walled nanotubes, scrolls offer superior performance for gas storage[29,30,31] and actuation[32,33,34], because they can easily radially expand and/or contract. Additionally, supercapacitor[35,36] and battery devices[35,37] benefit from the large and accessible interlayer surface area. The open-ended structure also adds a degree of freedom for impact deformations, as the scroll topology allows radial expansion/contraction in order to dissipate kinetic energy.

In particular, no studies detailing the CNS and BNS mechanical response to high strain-rate conditions have been carried out, and this is one of the objectives of the present work. We have investigated the dynamical and structural properties of CNS and BNS shot at high velocities against solid targets through fully atomistic MD simulations (see theoretical method section for details). We have considered conditions that mimic those used in the carbon[16] and boron nitride[17] nanotube shooting experiments. In addition, we have also used impact velocity values within the feasible experimental range.

Our results show that the critical parameters to determine impact outcome structures are the collision velocity and the relative scroll-target orientation. Both the orientation of the axis and of the open edge (see Figure 2) are important. In comparison to nanotubes, we observed a larger variety of resulting structures in nanoscroll collisions, due to the additional degrees of freedom for structural deformation.

## Theoretical Method

Structural, dynamical and mechanical properties of CNS and BNS at high velocity impact were investigated through fully atomistic reactive molecular dynamics (MD) simulations. We employed the ReaxFF force field[38,39], which is a reactive force field capable of reproducing the formation and/or breaking of chemical bonds in carbon and boron nitride-based structures. This potential is implemented in the cross-platform and open source code LAMMPS[40] and works as a bridge between quantum and classical methods, where the parameters are obtained directly from first principles calculations and/or experiments[38]. One of its main advantages is its relatively low computational cost, which allows us to handle large systems.

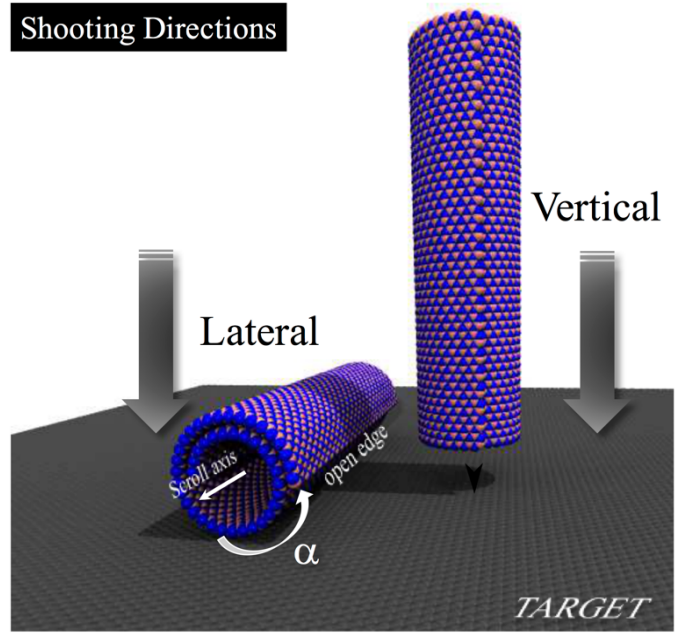

Figure 2: Nanoscroll impact orientations. Regarding the scroll axis orientation relative to the target, we considered lateral and vertical configurations. For lateral impacts, we also considered different α angle values.

In our MD simulations, we used 100 Å long CNS and BNS, with outer diameter values of ~25 Å (~4500 atoms). We restricted our study to defect-free, fully rolled-up nanoscrolls. We also tested chirality dependence, but for the impact velocity values considered here, our main conclusions are chirality independent. Due to the used high velocity and kinetic energy values, we used very small time-steps (0.025 fs) to avoid spurious numerical effects. In all our MD simulations, we kept the number of particles (N), volume (V), and energy (E) constant (NVE ensemble).

The high strain-rate regime was achieved by shooting the nanoscrolls at ultrasonic velocities against a fixed van der Waals wall (rigid substrate). For CNS, we considered shooting velocity values varying from 2 km/s to 6 km/s, with increment values of 0.5 km/s. For BNS, we considered values ranging from 2 km/s to 4 km/s (beyond this velocity the structures are destroyed, because BN is very brittle), with increment values of 0.4 km/s. In addition to different velocities, we also considered different relative orientations between nanoscrolls and targets. Figure 2 shows the shooting directions investigated: lateral and vertical ones. In the vertical case, we considered normal angle collisions of CNS/BNS against substrates. In the lateral case, a CNS/BNS collided with the substrate parallel to its axis. For this case, we also considered three values for the angle α displayed in Figure 2: $0°$, $90°$, and $180°$.

## Results

In Figure 3 we present representative MD snapshots for the lateral shooting cases. We grouped the resulting structures after impact into five general categories: unscrolled, semi-unscrolled, welded, amorphous, and torn. For the unscrolled case, the structure unravels into a two-dimensional sheet, with no significant damage to the underlying $sp^2$ atomic network. This configuration was observed only for CNS. For the semi-unscrolled case, there is a partial unravelling of the initial configuration, with damage to the underlying $sp^2$ network. For the welded case, reconstructions occur at impact, and inner and outer scroll layers become covalently bonded. For carbon scrolls with α=$0°$, in some cases reconstructions were limited to fusion of the open edge to an adjacent wall layer. The result was a novel nanostructure, presenting a scroll topology without an external open edge. In the amorphous case, inter-layer bonds are again formed at impact, but now there is substantial damage to the honeycomb network, particularly at the side that first collided with the target. Finally, in the torn case, there is massive atom ejection, with complete destruction of the scroll structural integrity. As we will discuss next, impact velocity and orientation determine which structure will be obtained from a shooting test. Some of these configurations were also observed for CNT shootings[16].

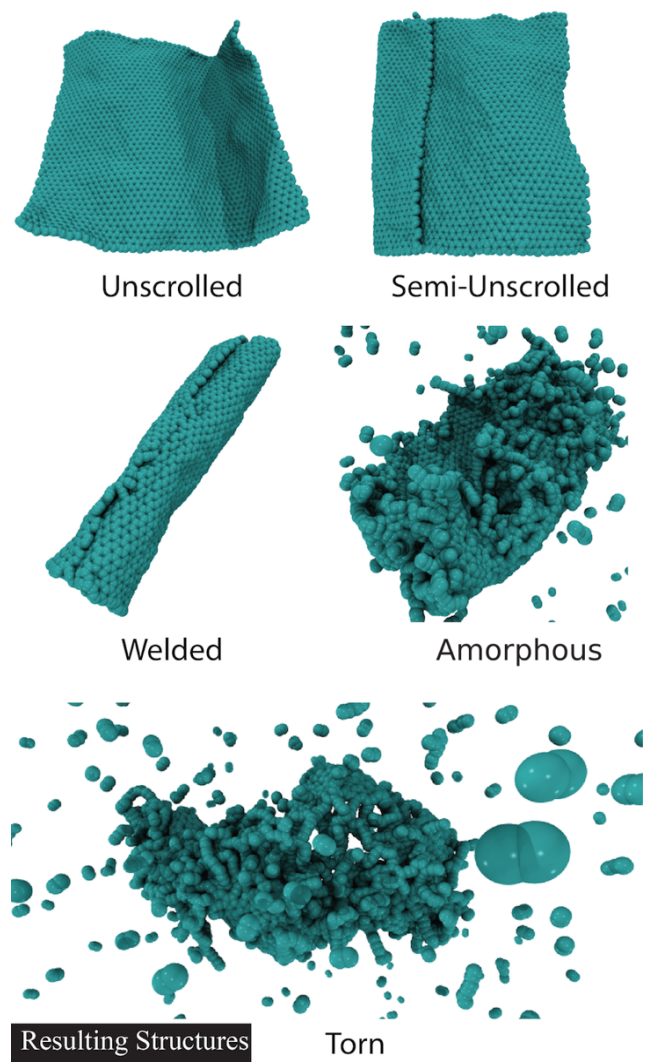

Figure 1: Representative MD snapshots of the resulting structures after lateral scroll impacts.

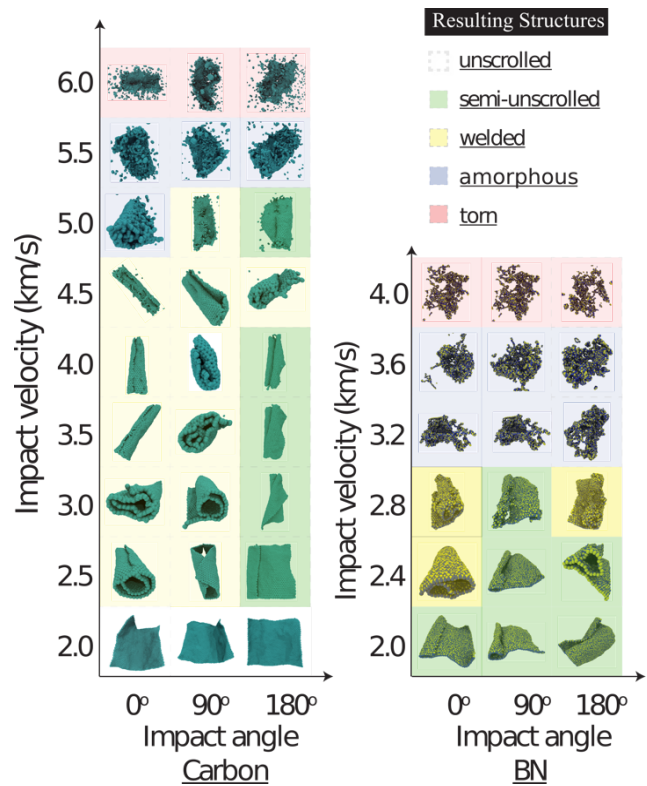

Figure 4: MD snapshots of the resulting structures after lateral scroll impacts, as a function of impact angle and velocity. Notice that snapshot orientation is unrelated to impact angle. For instance, many structures for α = $0°$ are depicted with their open edge upwards, but in the simulations they were facing the target (downwards). We observed less bond reconstruction at impact when BN based nanoscrolls were employed, due to their brittle nature.

A summary of results for the lateral impact cases, as a function of velocity and α angle values, is presented in Figure 5. For CNS, small shooting velocities (2.0 km/s) favoured unscrolled structures. In this case, the impact energy conversion (translational kinetic energy into potential and internal kinetic energy) does not induce reconstructions of the $sp^2$ network. At this lower velocity value, the increase in potential energy is merely enough to overcome the weaker van der Waals interactions, which maintained the graphene sheets scrolled. For BNS in the same velocity range, the stronger interlayer mechanical coupling[7] prevented the complete unravelling of the scrolls. Consequently, semi-unscrolled and welded structures were the observed configurations for impacts at lower velocities (<3.0 km/s). For CNS, unscrolled and welded configurations were also observed for impacts in the 2.5-4.5 km/s velocity range. A uniqueness of CNS is that partial unscrolling was only observed for shootings with α = 180°. For α = 0°, in this velocity range we observed the aforementioned close-ended nanoscrolls. Note that this material probably presents the same conducting characteristics reported for CNS[25], as every layer is still covalently connected. It is, however, unlikely that this structure presents the reported giant electroactuation[32], as it requires the presence of an open outer edge. Therefore, we expect a hybrid electromechanical response for these welded architectures, between nanoscroll and nanotube. For other α impact angles, welded structures still present open edges. Nevertheless, large radial expansion might also be prevented, as interlayer bonds were formed at impact.

For increased shooting velocities (>5 km/s for carbon and > 3.2 km/s for BN), we observed significant structural damage to the layers into direct contact with the target, in addition to bond reconstruction at other layers. This process produced amorphous structures. For even higher velocities, we observed the destruction of the honeycomb network of the rolled-up graphene/BN sheet, as well as, extensive atom ejections at impact. In the resulting torn structures, a fraction as low as 10% of the total atoms retained their original $sp^2$ hybridization. Similar configurations were also observed for CNT shootings[16]. For the case of BNS, it is important to emphasize that we obtained torn structures at lower velocities because the rolled-up BN sheets are very brittle.

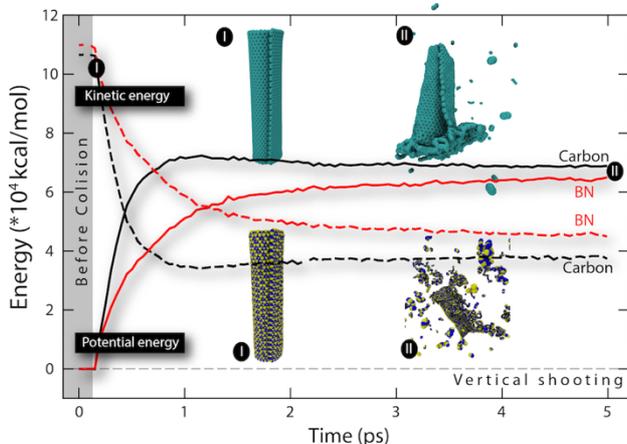

Figure 5: Potential and Kinetic energies as a function of simulation time for CNS (black curves) and BNS (red curves), for a vertical impact at a velocity of 4 km/s. "I" and "II" show, respectively, structures before and after impact. Part of the kinetic energy is converted into potential at impact, at different rates for the two materials.

We observed less variety of resulting structures for the vertical shooting cases. Typically, scroll regions close and/or into direct contact to the target are fractured, regions far from the target remained unbroken, and intermediate regions underwent structural reconstructions. In Figure 5, we show the kinetic energy (dashed lines) and potential energy (solid lines) as a function of simulation time, for a BNS (red lines) and CNS (black lines). For the BN-based material, many high velocity fragments were ejected at impact, resulting in less kinetic-to-potential energy conversion. This can be inferred from the MD snapshots presented as insets in Figure 5. For the h-BN scrolls, notice the large number of fragments surrounding the highly deformed BNS after impact. For the carbon material, the number of fragments is much smaller, and there is an amorphous region formed near the lower end. Quantitatively, for this velocity, we determined the carbon and BN mass loss to be 5.2% and 34.6%, respectively. Mass loss percentages for different velocities, materials, and orientations are provided in the electronic supplementary information (ESI). A better understanding of the whole processes can be obtained from the videos in the (ESI).

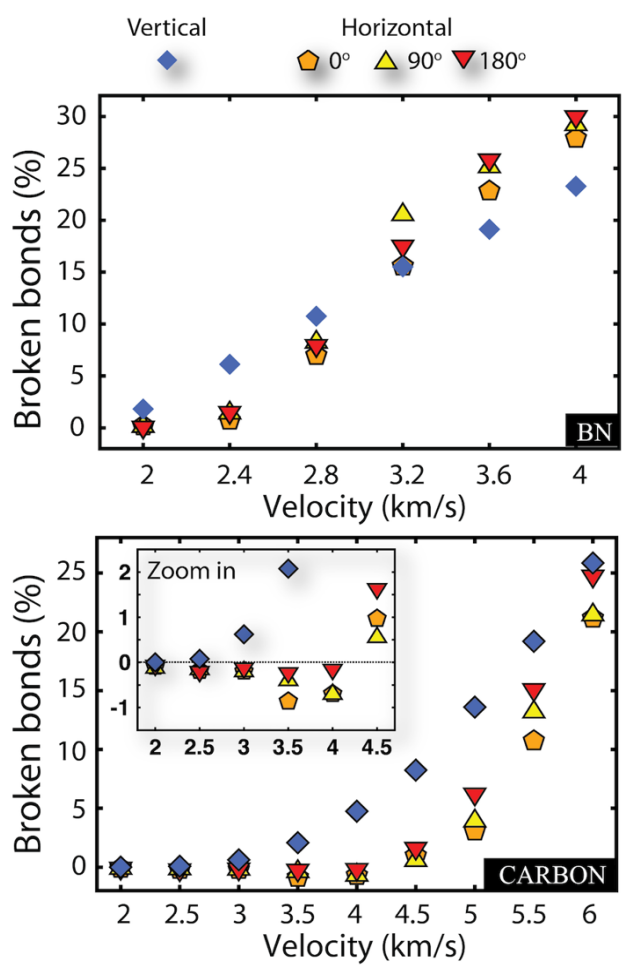

Figure 6: Percentage of broken bonds as a function of impact velocity values for BNS and CNS, respectively. The inset for the CNS case highlights the increase in the number of bonds (negative percentages, as new bonds are formed). The fraction of broken bonds increased almost monotonically for BNS.

In Figure 6, we show percentages of broken bonds for different shooting velocities and impact angles. For BNS, this percentage increases almost monotonically for every orientation/velocity, but at different rates. We can see that vertical shootings induce more broken bonds if v < 3 km/s, but less broken bonds if v > 3 km/s. For this orientation, stress is concentrated in less atoms[16], leading to mechanical failure

(fracture) even at lower impact velocities. However, at higher velocities, stress concentration contributes to preserve the structural integrity of the ends far away from the substrate. Lateral impacts, on the other hand, distribute stress through a larger area[16], thus preventing large-scale fracture at lower velocities. At higher velocities, however, stress dispersion decreases the loss of mass, but causes widespread structural damage, as illustrated for BNS impacts with v = 4.0 km/s. For the vertical impacts, although mass ejection corresponded to 34.6 % of the initial mass, still 32.8% of the atoms retained three neighbours (60.3% of the atoms lost one neighbour, and 6.7% lost two neighbours). For a lateral shooting with $\alpha$ = 180°, the loss of mass was smaller (19.3%), but only 10.3% of the atoms retained three neighbours. Of the remaining atoms, 85.7% preserved two neighbours, but 3.9% held a single neighbour after impact. Similar dynamics were observed for carbon-based scrolls. For example, if we compare the percentage of broken bonds for vertical and lateral shootings, we observe that the differences between the two values increased for lower impact velocities (v < 5.0 km/s) and then decreased for higher velocities (v > 5.0 km/s). Still, for the tested velocities, the amount of broken bonds was always higher for vertical shootings. A uniqueness of lateral shootings for CNS is that the fraction of broken bonds can be negative for lower impact velocities (< 4.0 km/s) – see the inset of Figure 6. When this occurred, the number of bonds increased during impact, a process that typically lead to the previously described welded structures.

## Conclusions

We carried out fully atomistic reactive molecular dynamics (MD) simulations to investigate the response of carbon and BN nanoscrolls to high strain-rate conditions, which were generated by shooting these structures at high velocities against solid targets. We considered different impact velocity values, as well as varied scroll axis and open edge orientations. Our MD results show these variables played a fundamental role in determining the resulting structures after impact. During impact, there is fast conversion of kinetic into potential energy. For lateral shooting cases, if the initial kinetic energy is below a threshold, this leads to partial or complete unscrolling, with the overcoming of weaker van der Waals interactions. On the other hand, for higher kinetic energies, the increase in potential energy results in extensive fracture of covalent bonds. For intermediate velocities, the number of bonds may actually increase, due to the formation of crosslinks between scroll layers. If the outer open edge is facing towards the substrate during impact, reconstruction might be limited to neighbouring impact regions, leading to a new kind of hybrid nanoscroll-nanotube close-ended nanostructure. Some of these configurations were not observed for BNS, due to the boron nitride brittle nature. For normal angle collisions, we observed less variety of resulting structures. We observed structural failures (fractures) in these shootings even at lower velocities, due to the concentration of stress into a smaller impact area. Interestingly, when comparing BNS lateral and vertical shootings, we observed that stress concentration prevented the fracture of areas far from the impact region. As a result, normal angle collisions actually produced less broken bonds at higher shooting velocities, a behaviour completely distinct to that of CNS.

Considering that CNS and BNS were already synthesized and the renewed interest in nanoscrolls[41-43] we hope the present study will stimulate experiments for nanoscrolls, similar to those already carried out for nanotubes[16-18].

## Acknowledgements

The CFW thanks São Paulo Research Foundation (FAPESP) Grant No. 2014/24547-1 for financial support. JMS thanks the support from Coordenação de Aperfeiçoamento de Pessoal de Nível Superior (CAPES) through the Science Without Borders program (Project Number A085/2013) and PNPD CAPES fellowship. Computational and financial support from the Center for Computational Engineering and Sciences at Unicamp through the FAPESP/CEPID Grant No. 2013/08293-7, is also acknowledged.

## Notes and references